# Deep Learning System to Screen Coronavirus Disease 2019 Pneumonia

Xiaowei Xu[1], MD; Xiangao Jiang[2], MD, Chunlian Ma[3], MD; Peng Du[4]; Xukun Li[4]; Shuangzhi Lv[5], MD; Liang Yu[1], MD; Yanfei Chen[1], MD; Junwei Su[1], MD; Guanjing Lang[1], MD; Yongtao Li[1], MD; Hong Zhao[1], MD; Kaijin Xu[1], PhD MD; Lingxiang Ruan[5], MD; Wei Wu[1], PhD MD

1. *State Key Laboratory for Diagnosis and Treatment of Infectious Diseases, National Clinical Research Center for Infectious Diseases, Collaborative Innovation Center for Diagnosis and Treatment of Infectious Diseases, the First Affiliated Hospital, Zhejiang University, Hangzhou 310003, China.*
2. *Department of Infectious Disease, Wenzhou central Hospital, Wenzhou, Zhejiang 325000, People's Republic of China*
3. *Department of Infectious Disease, First People's Hospital of Wenling, WenLing, Zhejiang 317500, People's Republic of China*
4. *Artificial Intelligence Lab, Hangzhou AiSmart-Image-Cloud-Analysis Co., Ltd., Hangzhou, Zhejiang 310012, People's Republic of China*
5. *Department of Radiology The First Affiliated Hospital, College of Medicine, Zhejiang University, Hangzhou, Zhejiang 310003, People's Republic of China*

Correspondence to Wei Wu, PhD, MD

Tel: +86 571 87236458; Fax: +86 571 87236459

E-mail: 1198042@zju.edu.cn

**Key Points**

**QUESTIONS:** Can artificial intelligence technology used to early screen COVID-19 patients from their computed tomography (CT) images and what is the diagnostic accuracy from computer?

**FINDINGS**: In this multi-center case study, the overall accuracy of the deep learning models were 86.7% for three groups: COVID-19, Influenza-A viral pneumonia and healthy cases.

**MEANING**: It is fully automatic and could be a promising supplementary diagnostic method for frontline clinical doctors.

**Abstract**

**IMPORTANCE:** We found that the real time reverse transcription-polymerase chain reaction (RT-PCR) detection of viral RNA from sputum or nasopharyngeal swab has a relatively low positive rate in the early stage to determine COVID-19 (named by the World Health Organization). The manifestations of computed tomography (CT) imaging of COVID-19 had their own characteristics, which are different from other types of viral pneumonia, such as Influenza-A viral pneumonia. Therefore, clinical doctors call for another early diagnostic criteria for this new type of pneumonia as soon as possible.

**OBJECTIVE:** This study aimed to establish an early screening model to distinguish COVID-19 pneumonia from Influenza-A viral pneumonia and healthy cases with pulmonary CT images using deep learning techniques.

**DESIGN:** The candidate infection regions were first segmented out using a 3-dimensional deep learning model from pulmonary CT image set. These separated images were then categorized into COVID-19, Influenza-A viral pneumonia and irrelevant to infection groups, together with the corresponding confidence scores using a location-attention classification model. Finally the infection type and total confidence score of this CT case were calculated with Noisy-or Bayesian function.

**SETTING:** CT samples contributed from three COVID-19 designated hospitals from Zhejiang Province, China.

**PARTICIPANTS:** A total of 618 CT samples were collected: 219 from 110 patients with COVID-19, 224 CT samples from 224 patients with Influenza-A viral pneumonia, and 175 CT samples from healthy people.

**RESULTS:** The experiments result of benchmark dataset showed that the overall accuracy was 86.7 % from the perspective of CT cases as a whole.

**CONCLUSIONS:** The deep learning models established in this study were effective for the early screening of COVID-19 patients and demonstrated to be a promising supplementary diagnostic method for frontline clinical doctors.

## 1. Introduction

At the end of 2019, the novel coronavirus disease 2019 pneumonia (COVID-19) occurred in the city of Wuhan, China.[1-4] On January 24, 2020, Huang et al.[5] summarized the clinical characteristics of 41 patients with COVID-19, indicating that the common onset symptoms were fever, cough, myalgia, or fatigue. All these 41 patients had pneumonia and their chest CT examination showed abnormalities. The complications included acute respiratory distress syndrome, acute heart injury, and secondary infections. Thirteen (32%) patients were admitted to the intensive care unit (ICU), and six (15%) died. The the Kok-KH[6] team at the University of Hong Kong found the evidence of human-to-human transmission of COVID-19 for the first time.

COVID-19 causes severe respiratory symptoms and is associated with relatively high ICU admission and mortality. The current clinical experience for treating these patients revealed that the RT-PCR detection of viral RNA from sputum or nasopharyngeal swab had a low positive rate in the early stage. However, a high proportion of abnormal chest CT images were obtained from patients with this disease. The manifestations of CT imaging of COVID-19 cases had their own characteristics, different from the manifestations of CT imaging of other viral pneumonia such as Influenza-A viral pneumonia, as showed in Figure 1. Therefore, clinical doctors called for replacing nucleic acid testing with lung CT as one of the early diagnostic criteria for this new type of pneumonia as soon as possible.

With the rapid development of computer technology, digital image processing technology has been widely applied in the medical field, including organ segmentation and image enhancement and repair, providing support for subsequent medical diagnosis.[7,8] Deep learning technologies, such as convolutional neural network (CNN) with the strong ability of nonlinear modeling, have extensive applications in medical image processing as well.[9-12] Relevant studies were conducted on the diagnosis of pulmonary nodules,[13-14] classification of benign and malignant tumors,[15] and pulmonary tuberculosis analysis and disease prediction[16-18] worldwide.

In this study, multiple CNN models were used to classify CT image datasets and calculate the infection probability of COVID-19. The findings might greatly assist in the early screening of patients with COVID-19.

## 2. Method

### *2.1 Dataset introduction*

A total of 618 transverse-section CT samples were collected in this study, including 219 from 110 patients with COVID-19 from the First Affiliated Hospital of Zhejiang University, the No.6 People's Hospital of Wenzhou, and the No.1 People's Hospital of Wenling , from Jan 19 to Feb 14, 2020. All three hospitals are designated COVID-19 hospitals in Zhejiang Province. Every COVID-19 patient was confirmed with RT-PCR testing kit and we also excluded the cases that had no image manifestations on the chest CT images. In addition, there had at least two days gap between CT datasets if taken from the same patient to ensure the diversity of samples. The remaining 399 CT samples were collected from the First Affiliated Hospital of Zhejiang University as the controlled experiment group. Among them, 224 CT samples were from 224 patients with Influenza-A viral pneumonia including H1N1, H3N2, H5N1, H7N9 etc., and 175 CT samples from healthy people. There were 198 (90.4%) COVID-19 and (196) 86.6% Influenza-A cases from early or progressive stages and the rest 9.6% and 13.4% cases from severe stage respectively ($P > 0.05$). Moreover, Influenza-A viral pneumonia CT cases used as it was most critically to distinguish them from suspected patients with COVID-19 currently in China.

The ethics committee of the First Affiliated Hospital, College of Medicine, Zhejiang University approved this study and all research was performed in accordance with relevant guidelines/regulations. All participants and/or their legal guardians signed the informed consent form before the study.

A total of 528 CT samples (85.4%) were used for training and validation sets, including 189 samples of patients with COVID-19, 194 samples from patients with Influenza-A viral pneumonia, and 145 samples from healthy people. The remaining 90 CT sets (14.6%) were used as the test set, including 30 COVID-19, 30 Influenza-A viral pneumonia, and 30 healthy case. Furthermore, the test cases of CT set were selected from the people who had not been included in the training stage.

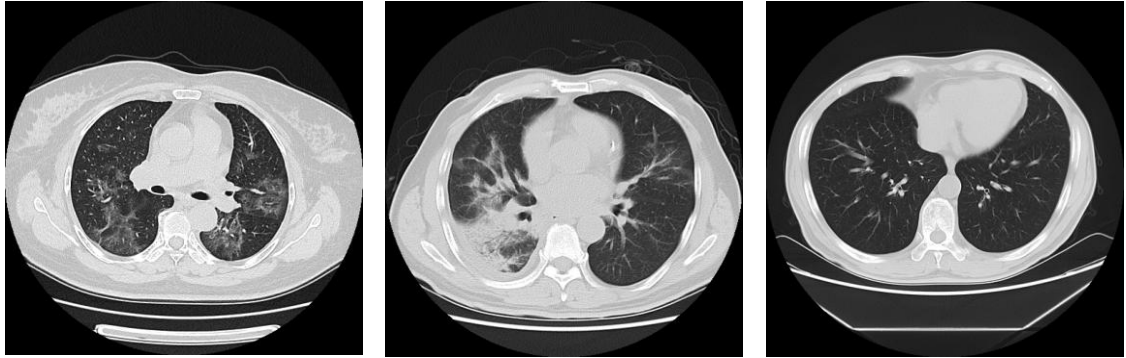

(a) (b) (c)

**Figure 1.** Typical transverse-section CT images: (a) COVID-19; (b) Influenza-A viral pneumonia; (c) no pneumonia manifestations on this chest CT image.

## 2.2 Process

Figure 2 shows the whole process of COVID-19 diagnostic report generation in this study. First, the CT images were preprocessed to extract effective pulmonary regions. Second, a 3D CNN model was used to segment multiple candidate image cubes. The center image, together with the two neighbors of each cube, was collected for further steps. Third, an image classification model was used to categorize all the image patches into three types: COVID-19, Influenza-A-viral-pneumonia, and irrelevant-to-infection. Image patches from the same cube voted for the type and confidence score of the candidate as a whole. Finally, the overall analysis report for one CT sample was calculated using the Noisy-or Bayesian function.[19]

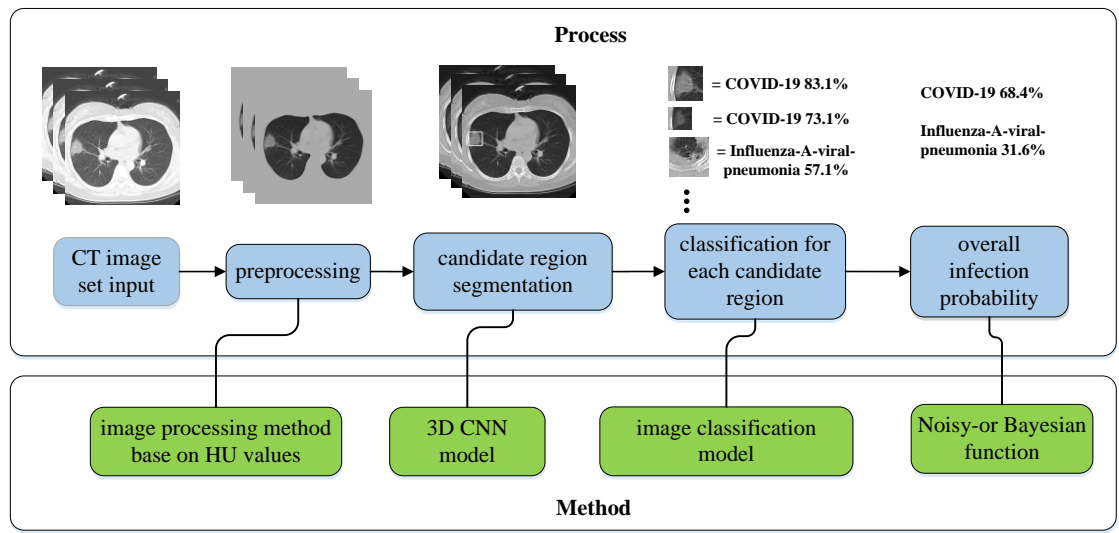

**Figure 2**. Process flow chart.

## 2.3 Dataset preprocessing and candidate region segmentation

To study was expedited using the same method and models in data preprocessing and candidate region segmentation stages as a previous study on pulmonary tuberculosis.[17] The focus of infections from pulmonary tuberculosis had multiple structures and types, including miliary, infiltrative, caseous, tuberculoma, and cavitary etc. Although, the VNET[20] based segmentation model VNET-IR-RPN[17] was trained for pulmonary tuberculosis purpose, it was verified to be still good enough to separate candidate patches from viral pneumonia.

Moreover, in the study of pulmonary tuberculosis, the VNET-IR-RPN model was used both for both segmentation and classification. Only the segmentation-related bounding box regression part was preserved, regardless of the classification results, because only the former task was required at this stage in this study.

*2.4 Image data processing and augmentation*

A large number of non-infection regions irrelevant to this study were also separated using the 3D segmentation model, including fibrotic structure of pulmonary, calcification spots, or healthy regions identified incorrectly. Therefore, an extra category was added as irrelevant-to-infection besides COVID-19 and Influenza-A-viral-pneumonia.

The study included 618 CT samples (219 COVID-19, 224 Influenza-A-viral-pneumonia, and 175 healthy case). Subsequently, a total of 3957 candidate cubes were generated from the 3D segmentation model. Only the territory close to the middle of this cube contained maximum information about this focus of infection. Hence, only the center image together with the two neighbors of each cube was collected to represent this region for future classification steps. Next, all image patches were manually classified by two professional radiologists into two types: irrelevant-to-infection and pneumonia. The the images in the latter category were recognized automatically as COVID-19 or Influenza-A-viral-pneumonia based on the clinical diagnosis results.

A total of 11,871 image patches were acquired from the aforementioned steps, including 2,634 COVID-19, 2,661 Influenza-A-viral-pneumonia, and 6,576 irrelevant-to-infection. According to the previous dataset assignment, the training and validation sets had 528 CT samples, equivalent to 10,161 (85.6%) images, including 2,301 COVID-19, 2,244 Influenza-A-viral-pneumonia, and 5,616 irrelevant-to-infection images. The remaining 1,710 (14.4%) images were reserved for the test dataset.

The sampling possibility of COVID-19 and Influenza-A-viral-pneumonia cases was expanded three times to balance the specimen number of irrelevant-to-infection so as to reduce the influence of the uneven distribution of different image types on the present dataset. At the same time, generic data expansion mechanisms, such as random clipping, left-right, up-down flipping, and mirroring operation, were performed on specimens to increase the number of training samples and prevent data overfitting.

## 2.5 Deep learning model for classification

### 2.5.1 Location-attention classification

The work of Jeffrey Kanne[21] and Chung M et al,[22] showed at least three distinguishing features of COVID-19: ground-glass appearance, striking peripheral distribution along with the pleura, and usually more than one independent focus of infections for one case, as showed in Figure 3.

The models were optimized based on these findings. The image classification model was designed to distinguish the appearance and structure of different infections. Moreover, relative distance-from-edge as an extra weights, was used for the model to learn the relative location information of the patch on the pulmonary image. The focus of infections located close to the pleura were be more likely to be recognized as COVID-19.

The relative distance-from-edge of each patch was calculated as followings:

1) Measure the minimum distance from the mask to the center of this patch (double-headed arrow as showed in Fig 3c)

2) Obtain the diagonal of the minimum circumscribed rectangle of the pulmonary image (Fig 3d)

3) Then the relative distance-from-edge achieved by the distance obtained from step 1) divided by the diagonal from step 2).

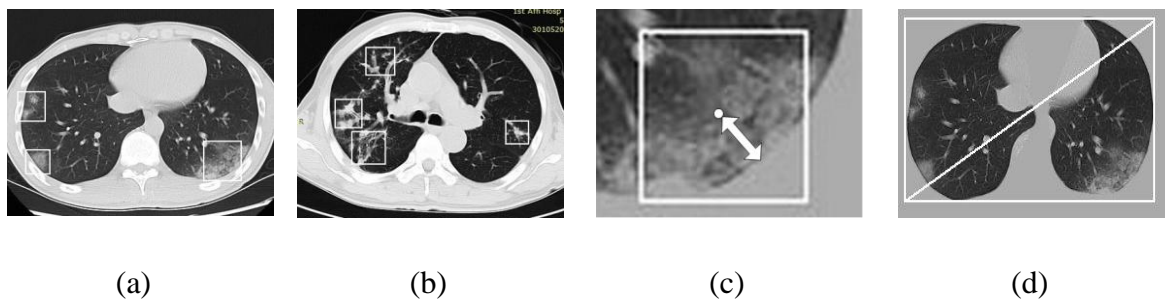

(a)          (b)          (c)          (d)

**Figure 3**. (a) COVID-19 image with three ground-glass focus of infections; (b) Influenza-A-viral-pneumonia image with four focus of infections; (c) the minimum distance from the mask to the center of this patch (double-headed arrow) and (d) diagonal of the minimum circumscribed rectangle of this pulmonary image.

*2.5.2 Network structure*

Two CNN three-dimensional classification models were evaluated in this study, as showed in Figure 4. One was a relative traditional ResNet[23]-based network and another model was designed based on the first network structure by concatenating the location-attention mechanism in the full-connection layer to improve the overall accuracy rate.

The classical ResNet-18 network structure was used for image feature extraction. Pooling operations were also used for the dimensional reduction of data to prevent overfitting and improve the problem of generalization.

The output of the convolution layer was flattened to a 256-dimensional feature vector and then converted into a 16-dimensional feature vector using a full-connection network. For the location-attention network, the value of relative distance-from-edge was first normalized to the same order of magnitude and then concatenated to this full-connection network structure. Next, three full-connection layers were followed to output the final classification result together with the confidence score.

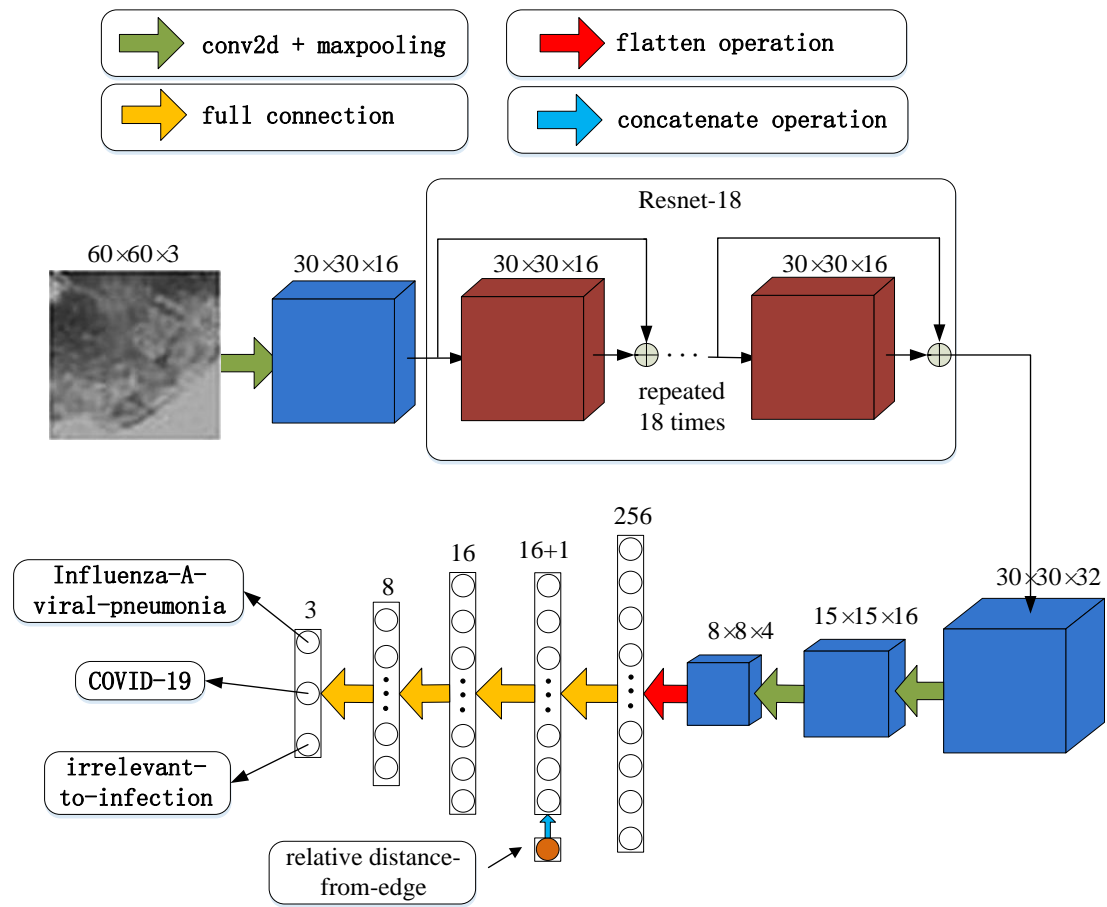

**Figure 4**. Network structure of the location-attention oriented model.

*2.6 Diagnostic report*

*2.6.1 Vote for each candidate region*

Inspired by the theory of Bagging prediction algorithm[24] in machine learning technology, one candidate region was represented by three image patches: the center image together with its two neighbors. These three images voted for this whole region with the following strategies:

1) If at least two images were categorized into the same type, then the image with the maximum confidence score in this type was selected

2) Otherwise, the image with the maximum confidence score was picked (no type dominated).

Regions that voted as irrelevant-to-infection type was ignored in the next step.

*2.6.2 Noisy-or Bayesian function to deduce the overall report*

One of the remarkable features of COVID-19 is more than one independent focus of infections in one CT case. It is reasonable that the overall probability is much larger than 50% if a patient has two COVID-19 regions, both having a 50% probability. Accordingly, the total infection confidence score (*P*) for one infection type was calculated using the probability formula of the Noisy-or Bayesian function as follows:

$$P = 1 - \prod_i (1 - P_i) \tag{1}$$

where $P_i$ represents the confidence of the *i*th region.

The confidence scores of two types $P_{COVID-19}$ and $P_{Influenza-A-viral-pneumonia}$ were deduced accordingly, then this CT sample was categorized to the corresponding group according to the dominated $P$ value.

Moreover, the following strategies were used output the confidence possibility of an entire CT sample for a reasonable reference to clinical doctors:

1. If both $P$ values were equal to 0, then this CT sample belonged to the no-infection-found case

2. If one of the $P$ values was equal to 0, then the other $P$ value was exported directly as the confidence possibility of this CT sample

3. Otherwise, the softmax function was used to generate two confidence scores.

$$S_i = \frac{e^{P_i}}{\sum_j e^{P_j}} \qquad (2)$$

where $i, j \in (COVID-19, Influenza-A-viral-pneumonia)$. Both $S_i$ was exported as the confidence score for each type of infection. The softmax operation normalized the sum of $S_i$ to 100% and did not alter the judgment result of infection types.

## 3. Results

### 3.1 Evaluation platform

An Intel i7-8700k CPU together with NVIDIA GPU GeForce GTX 1080ti was used as the testing server. The processing time largely depended on the number of image layers in one CT set. On average, it was less than 30s for a CT set with 70 layers from data-preprocessing to the output of the report.

## 3.2 Training process

As one of the most classical loss function used in classification models, cross-entropy was used in this study. When the epoch number of training iterations increased to more than 1000, the loss value did not decrease or increase obviously, suggesting that the models converged well to a relative optimal state without distinct overfitting. The training curves of the loss value and the accuracy for two classification models were showed in Figure 5. The network with the location-attention mechanism achieved better performance on the training dataset compared with the original ResNet.

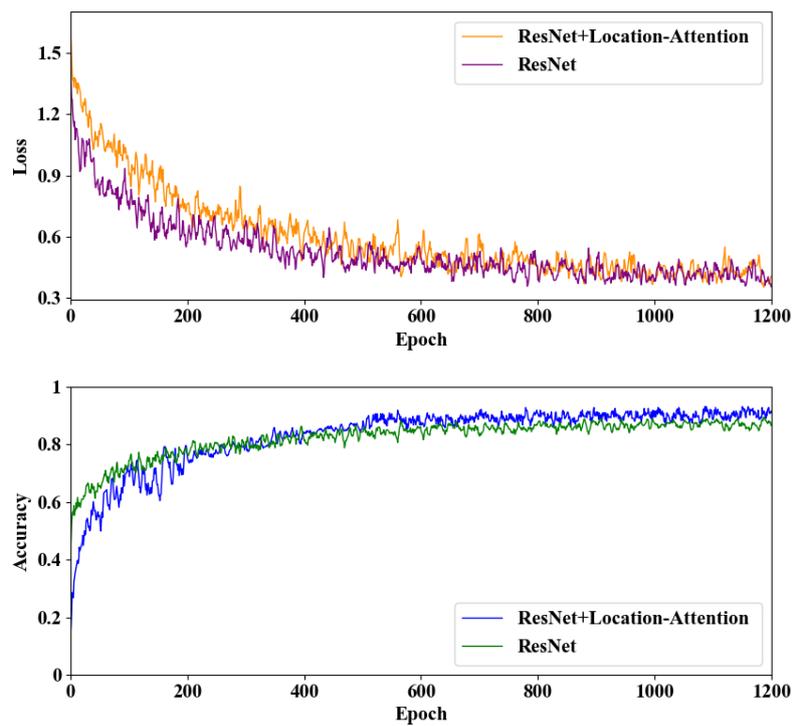

**Figure 5**. Training curve of loss and accuracy for the two classification models.

## 3.3 Performance on test dataset

### 3.3.1 Performance measurement

The *accuracy* of a method determines how correct the values are predicted. *Precision* determines the reproducibility of the measurement or how many of the predictions are correct. *Recall* shows how many of the correct results are discovered. *F1-score* uses a combination of precision and recall to calculate a balanced average result. The following equations show how to calculate these values, where TP, TN, FP and FN are true positive, true negative, false positive, and false negative respectively.

$$accuracy = \frac{TP+TN}{TP+FP+TN+FN} \quad (3)$$

$$precision = \frac{TP}{TP+FP} \quad (4)$$

$$recall = \frac{TP}{TP+FN} \quad (5)$$

$$f1-score = \frac{2 \times precision \times recall}{precision + recall} \quad (6)$$

### 3.3.2 Segmentation

A total of 30 CT samples were selected randomly from each group (COVID-19, Influenza-A-viral-pneumonia and healthy) for the test set, following the rule that this person (owner of this CT) had not been included in the previous training stage. Moreover, the segmentation model VNET-IR-RPN was configured to reduce the proposal's threshold to maximum separate candidate regions even through many normal regions could be included in. One CT case from COVID-19 group that had no region segmented as COVID-19 or Influenza-A-viral-pneumonia and hence wrongly categorized into the no-infection-found group, as showed in Figure 6. These focus of infections could be barely notified by human eyed and seemed too tenuous to be captured by the segmentation model for this study.

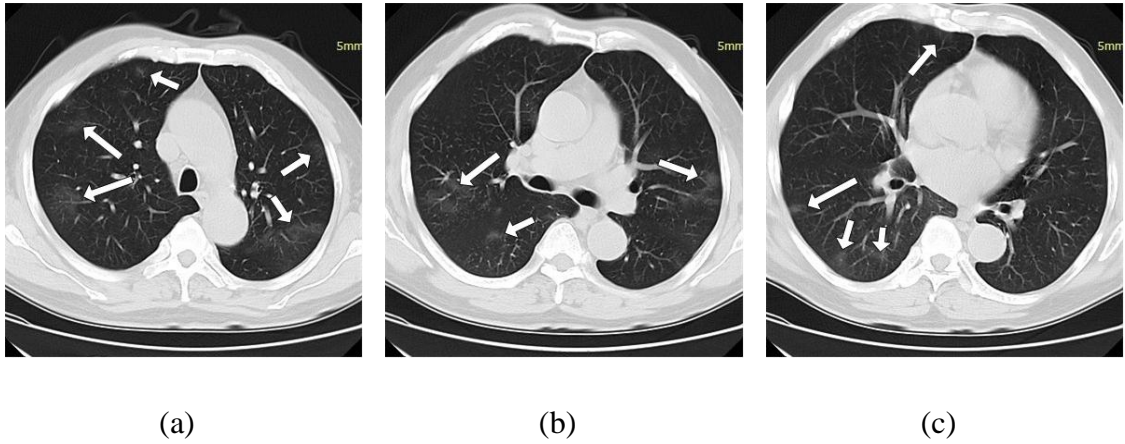

(a) (b) (c)

**Figure 6**. All CT images from a single CT case. The focus of infections are pointed out by arrows.

*3.3.3 Classification for a single image patch*

A total of 1,710 image patches were acquired from 90 CT samples, including 357 COVID-19, 390 Influenza-A-viral-pneumonia, and 963 irrelevant-to-infection (ground truth). To determine which was the optimal approach, the design of each methodology was assessed using a confusion matrix. Two network structures were evaluated: with and without the location-attention mechanism, as showed in Table 1 and Table 2.

|  |  | Predicted result | | |
|---|---|---|---|---|
|  |  | COVID-19 ($M_1/M_2$) | IAVP ($M_1/M_2$) | ITI ($M_1/M_2$) |
| Actual result | COVID-19 ($M_1/M_2$) | 260/273 | 47/32 | 50/52 |
|  | IAVP ($M_1/M_2$) | 55/46 | 276/280 | 59/64 |
|  | ITI ($M_1/M_2$) | 75/77 | 81/82 | 807/804 |

**Table 1.** The confusion matrix of COVID-19, Influenza-A-viral-pneumonia (IAVP) and irrelevant-to-infection (ITI). M1 and M2 referred to ResNet model and ResNet with location-attention mechanism model.

|  | Recall ($M_1,M_2$) | Precision ($M_1,M_2$) | f1-score ($M_1,M_2$) |
| --- | --- | --- | --- |
| COVID-19 ($M_1,M_2$) | 0.728/0.765 | 0.667/0.689 | 0.696/0.725 |
| IAVP ($M_1,M_2$) | 0.708/0.718 | 0.683/0.711 | 0.695/0.714 |
| ITI ($M_1,M_2$) | 0.838/0.835 | 0.881/0.874 | 0.859/0.854 |

**Table 2.** The recall, precision and f1-score of two classification models for COVID-19, Influenza-A-viral-pneumonia (IAVP) and irrelevant-to-infection (ITI). M1 and M2 referred to ResNet model and ResNet with location-attention mechanism model.

The average f1_scores for two models were 0.750 and 0.764, which indicated that the second model with location-attention mechanism achieved better performance averagely. Therefore, this model was used for the rest of this study.

*3.3.4 Vote for a region*

Each image patch voted to represent this entire candidate region. A total of 570 candidate cubes were distinguished, including 119 COVID-19, 130 Influenza-A-viral-pneumonia and 321 irrelevant-to-infection regions (ground truth). The confusion matrix of voting result and corresponding recall, precision and f1-score were showed in Table 3 and Table 4. The average f1-score for three categories was 0.856 and had an improvement of 12.1% compared with previous step.

|  |  | Predicted result | | |
|---|---|---|---|---|
|  |  | COVID-19 | IAVP | ITI |
| Actual result | COVID-19 | 97 | 15 | 7 |
|  | IAVP | 18 | 98 | 14 |
|  | ITI | 5 | 2 | 314 |

**Table 3.** The confusion matrix of COVID-19, Influenza-A-viral-pneumonia (IAVP) and irrelevant-to-infection (ITI).

|  |  | Recall | Precision | f1-score |
|---|---|---|---|---|
|  |  |  |  |  |
| Actual result | COVID-19 | 0.815 | 0.808 | 0.811 |
|  | IAVP | 0.754 | 0.852 | 0.800 |
|  | ITI | 0.978 | 0.937 | 0.957 |

**Table 4.** The recall, precision and f1-score of after voting process for each regions: COVID-19, Influenza-A-viral-pneumonia (IAVP) and irrelevant-to-infection (ITI).

*3.3.5 Result of the classification for the CT samples as a whole*

Noisy-or Bayesian function was used to identify the dominated infection types. Three types result exported in the final report: COVID-19, Influenza-A-viral-pneumonia and no-infection-found. The experimental results are summarize in Table 5 and Table 6. As the irrelevant-to-infection items (previous step) would be ignored and not be counted by the Bayesian function, we only compare the average f1-score for the first two items. They were 0.806 and 0.843 respectively, which showed a promotion of 4.7%. Moreover, the overall classification accuracy for all three groups are 86.7%.

|  | Predicted result | | |
|---|---|---|---|
|  | COVID-19 | IAVP | NIF |

|  |  | COVID-19 | IAVP | NIF |
|---|---|---|---|---|
| Actual result | COVID-19 | 26 | 3 | 1 |
|  | IAVP | 4 | 25 | 1 |
|  | NIF | 2 | 1 | 27 |

Table 5. The confusion matrix of COVID-19, Influenza-A-viral-pneumonia (IAVP) and no-infection-found (NIF).

|  |  | Recall | Precision | f1-score |
|---|---|---|---|---|
| Actual result | COVID-19 | 0.867 | 0.813 | 0.839 |
|  | IAVP | 0.833 | 0.862 | 0.847 |
|  | NIF | 0.900 | 0.931 | 0.915 |

Table 6. The recall, precision and f1-score from the output of the Bayesian function for COVID-19, Influenza-A-viral-pneumonia (IAVP) and no-infection-found (NIF).

## 4. Discussion

COVID-19, which was first detected in Wuhan China, has caused serious public health safety problems and hence become a global concern.[25-27] The severe situation puts forward new requirements for the prevention and control strategy. A large number of patients with viral pneumonia had been detected in Wuhan city. The RT-PCR test of 2019-nCoV RNA can make a definite diagnosis of COVID-19 from Influenza-A viral pneumonia patients. However, the nucleic acid testing has some defects, such as time lag, relatively low detection rate, and short of supply. In the early stage of COVID-19, some patients may already have positive pulmonary imaging findings but they have no sputum and negative test results in nasopharyngeal swabs of RT-PCR. These patients are not diagnosed as suspected or confirmed cases.

Thus, they are not be isolated or treated for the first time, making them potential sources of infection.

The CT imaging of COVID-19 present several distinct manifestations according to previous studies.[21,22] The manifestations include focal ground glass shadows mainly distributed in bilateral lungs, multiple consolidation shadows accompanied by the "halo sign" of surrounding ground glass shadow in both lungs, mesh shadows and bronchiectasis and inflating signs inside the lesions, and multiple consolidation of different sizes and grid-shaped high-density shadows. However, it is not objective and accurate to distinguish COVID-19 from other diseases only with human eyes. In comparison, deep learning system-based screen models revealed more specific and reliable results by digitizing and standardizing the image information. Hence, they can assist physicians to make a quick clinical decision more accurately, which would benefit on management of suspected patients.

In this study, the deep learning technology was used to design a classification network for distinguishing the COVID-19 from Influenza-A viral pneumonia. In terms of the network structure, the classical ResNet was used for feature extraction. It was compared with the network model with and without the added location-attention mechanism. The experiment showed that the aforementioned mechanism could better distinguish COVID-19 cases from others.

The manifestation of COVID-19 may have some overlap with the manifestations of other pneumonias such as Influenza-A viral pneumonia, organic pneumonia and eosinophilic pneumonia. The clinical diagnosis of COVID-19 needs to combine the patients' contact history, travel history, first symptoms and laboratory examination. In this study, the number of model samples was limited. Hence, the training and test the number of samples should be expand to improve the accuracy in the future. More multi-center clinical studies should be conducted to cope with the complex clinical situation.

Moreover, efforts should be made to improve the segmentation and classification model. A better exclusive models should be designed for training, the segmentation and classification accuracy of the model should be improved, and the generalization performance of this algorithm should be verified with a larger data set.

## 5. Conclusion

In this multi-center case study, we had presented a novel method that could screen COVID-19 fully automatically by deep learning technologies. Models with location-attention mechanism could more accurately classify COVID-19 at chest radiography with the overall accuracy rate of 86.7 % and could be a promising supplementary diagnostic method for frontline clinical doctors.

References

1. Zhu N, Zhang D, Wang W, et al. A Novel Coronavirus from Patients with Pneumonia in China, 2019[J]. N Engl J Med. 2020 Jan 24. doi: 10.1056/NEJMoa2001017.

2. Li Q, Guan X, Wu P, et al. Early Transmission Dynamics in Wuhan, China, of Novel Coronavirus-Infected Pneumonia[J]. N Engl J Med. 2020 Jan 29. doi: 10.1056/NEJMoa2001316.

3. Cohen J, Normile D. New SARS-like virus in China triggers alarm[J]. Science. 2020 Jan 17; 367(6475):234-235. doi: 10.1126/science.367.6475.234.

4. Corman VM, Landt O, Kaiser M, et al. Detection of 2019 novel coronavirus (2019-nCoV) by real-time RT-PCR[J]. Euro Surveill. 2020 Jan; 25(3). doi: 10.2807/1560-7917.ES.2020.25.3.2000045.

5. Huang C, Wang Y, Li X, et al. Clinical features of patients infected with 2019 novel coronavirus in Wuhan, China[J]. Lancet. 2020 Jan 24. pii: S0140-6736(20)30183-5. doi:10.1016/S0140-6736(20)30183-5.

6. Chan JF, Yuan S, Kok KH, et al. A familial cluster of pneumonia associated with the 2019 novel coronavirus indicating person-to-person transmission: a study of a family cluster[J]. Lancet. 2020 Jan 24. pii: S0140-6736(20)30154-9. doi: 10.1016/S0140-6736(20)30154-9.

7. Liu X, Guo S, Yang B, et al. Automatic Organ Segmentation for CT Scans Based on Super-Pixel and Convolutional Neural Networks[J]. Journal of Digital Imaging, 2018, 31(6).

8. Gharbi, Michaël, Chen J, Barron J T, et al. Deep Bilateral Learning for Real-Time Image Enhancement[J]. Acm Transactions on Graphics, 2017, 36(4):118.

9. Hesamian M H, Jia W, He X, et al. Deep Learning Techniques for Medical Image Segmentation: Achievements and Challenges[J]. Journal of Digital Imaging, 2019, 32(8).

10. Akagi M, Nakamura Y, Higaki T, et al. Correction to: Deep learning reconstruction improves image quality of abdominal ultra-high-resolution CT[J]. European Radiology, 2019, 29(8).

11. Nardelli P, Jimenez-Carretero D, Bermejo-Pelaez D, et al. Pulmonary Artery-Vein Classification in CT Images Using Deep Learning[J]. IEEE Transactions on Medical Imaging, 2018, PP(99):1-1.

12. Zhu W, Huang Y, Zeng L, et al. AnatomyNet: Deep learning for fast and fully automated whole‐volume segmentation of head and neck anatomy[J]. Medical Physics, 2019, 46(2).

13. Huang P, Park S, Yan R, et al. Added Value of Computer-aided CT Image Features for Early Lung Cancer Diagnosis with Small Pulmonary Nodules: A Matched Case-Control Study. Radiology. 2018 Jan; 286(1):286-295. doi: 10.1148/radiol.2017162725.

14. Ardila D, Kiraly AP, Bharadwaj S, et al. End-to-end lung cancer screening with three-dimensional deep learning on low-dose chest computed tomography. Nat Med. 2019 Jun;25(6):954-961. doi: 10.1038/s41591-019-0447-x.

15. Esteva A, Kuprel B, Novoa RA, et al. Dermatologist-level classification of skin

cancer with deep neural networks. Nature. 2017 Feb 2;542(7639):115-118. doi: 10.1038/nature21056. Epub 2017 Jan 25.

16. Lakhani P, Sundaram B. Deep Learning at Chest Radiography: Automated Classification of, Pulmonary Tuberculosis by Using Convolutional Neural Networks[J]. Radiology, 2017:162326.

17. Wu W, Li X, Du P, et al. A Deep Learning System That Generates Quantitative CT Reports for Diagnosing Pulmonary Tuberculosis. arXiv preprint arXiv:1910.02285, 2019.

18. Li L, Huang H, Jin X. AE-CNN Classification of Pulmonary Tuberculosis Based on CT Images[C]// 2018 9th International Conference on Information Technology in Medicine and Education (ITME). IEEE Computer Society, 2018.

19. Onisko A, Druzdzel M, Wasyluk H. Learning Bayesian network parameters from small data sets: application of Noisy-OR gates[J]. International Journal of Approximate Reasoning, 2001, 27(2):165-182.

20. Milletari F, Navab N, Ahmadi S A. V-Net: Fully Convolutional Neural Networks for Volumetric Medical Image Segmentation. arXiv preprint arXiv:1606.04797, 2016.

21. Jeffrey P Kanne. Chest CT Findings in 2019 Novel Coronavirus (2019-nCoV) Infections from Wuhan, China: Key Points for the Radiologist[J].Radiology, 2020.

22. Chung M, Bernheim A, Mei X, et al. CT Imaging Features of 2019 Novel Coronavirus (2019-nCoV) [J]. Radiology, 2020.

23. He K, Zhang X, Ren S, et al. Deep Residual Learning for Image Recognition[J].
**27** / **29**

2015.

24. Breiman, Leo. Bagging predictors[J]. Machine Learning, 24(2):123-140.

25. Wang C, Horby PW, Hayden FG, et al. A novel coronavirus outbreak of global health concern[J]. Lancet. 2020 Jan 24. pii: S0140-6736(20)30185-9. doi:10.1016/S0140-6736(20)30185-9.

26. Holshue ML, DeBolt C, Lindquist S, et al. First Case of 2019 Novel Coronavirus in the United States[J]. N Engl J Med. 2020 Jan 31. doi: 10.1056/NEJMoa2001191.

27. Chen N, Zhou M, Dong X, et al. Epidemiological and clinical characteristics of 99 cases of 2019 novel coronavirus pneumonia in Wuhan, China: a descriptive study[J]. Lancet. 2020 Jan 30. pii: S0140-6736(20)30211-7. doi:10.1016/S0140-6736(20)30211-7.

**Acknowledgments**

This study was supported by the China National Science and Technology Major Project Fund (20182X10101-001).

**Competing interests**

The authors declare no competing interests.

**Author contributions**

Wei Wu and Xiaowei Xu initiated the project and provided clinical expertise and guidance on the study design. Xukun Li and Peng Du desiged the network architecture and data/modeling infrastructure, training, testing setup and statistical analysis. Xiaowei Xu, Wei Wu, Xukun Li and Peng Du wrote the manuscript. Xiangao Jiang, Chunlian Ma, Shuangzhi Lv, Liang Yu, Yanfei Chen, Junwei Su, Guanjing Lang, Yongtao Li, Hong Zhao, Kaijin Xu and Lingxiang Ruan collected the datasets and interpreted the data. Xiaowei Xu, XianGao Jiang, Chun-Lian Ma and Peng Du contributed equally to this article.